# X-Ray Timing Properties of the Black Hole Candidate IGR J17091–3624 During the 2022 Outburst Onset with AstroSat

**Sree Bhattacherjee[1*], Biplob Sarkar[1]**

[1]Department of Applied Sciences, Tezpur University, Napaam, Assam 784028, India

*Email id: sbtezu@tezu.ernet.in (Corresponding Author)

**Abstract:** We present the evolution of rapid X-ray timing variability in the black hole X-ray binary IGR J17091-3624 during the initial phase of its 2022 outburst using a ~3-day-long AstroSat observation. Utilizing the high time-resolution capability of the Large Area X-ray Proportional Counter, we perform a comprehensive study of the evolution of Type-C quasi-periodic oscillations (QPOs) throughout the observation. The QPO centroid frequency evolves from 3.30 to 7.78 Hz, accompanied by the broad-band noise characteristics. We further investigate the energy dependence of the fractional rms variability and time lags, finding that the fractional rms exhibits a positive energy dependence, with its amplitude evolving throughout the observation, while the time lags exhibit the soft-lag behavior and evolve with QPO frequency. These findings provide important insights into the evolution of geometry of the inner accretion flow during the onset phase of the 2022 outburst of IGR J17091-3624.

## 1 Introduction

X-ray binaries (XRBs) are a type of accreting system that consists of an accretor, which could either be a black hole (BH) or a neutron star (NS) and a regular star moving around a common center of mass. The compact object accrete the hot and diffused matter from the companion star. The process through which it accretes depends on the type of XRB which is further categorised into low-mass XRBs (LMXBs) and high-mass XRBs (HMXBs) based on the mass of the companion star, which is ≤ 1 solar mass and ≥ 10 solar mass, respectively. XRBs are basically a potential candidate to investigate the accretion process in extreme physical environments and help understand the accretion flow dynamics and the geometry as well. In order to probe that, a comprehensive timing and spectral study is useful. In this work we focus on the study of the timing properties; there are two main types of timing variabilities in XRB systems, the broad-band noise and Quasi-periodic oscillations (QPOs). In BH-XRB systems, QPOs are broadly classified to low frequency QPOs (LFQPOs) (varying from mHz to few tens of Hz) and high frequency QPOs (HFQPO) (varying between few tens to few hundred Hz). LFQPOs observed in BH-XRBs are further classified into three classes: Type-A, B, and C, based on their centroid frequency ($\nu$), quality (Q) factor, fractional root-mean square (frms) amplitude, and associated spectral states (Remillard et al. 2002, Casella et al. 2005). Among these, Type-C QPOs are the most frequently observed and are commonly associated with the low-hard state (LHS) and hard-intermediate state (HIMS). They exhibit relatively high Q factors ($Q=\nu/FWHM$), $Q \gtrsim 5$, strong frms variability, and frequencies ranging from approximately 0.1–30 Hz and are superimposed on flat-top broad-band noise (Casella et al. 2005; Soleri et al. 2008). As the source evolves toward softer spectral states, the centroid frequency of the Type-C QPO generally increases (Motta et al. 2016). Even after decades of research, the origin of QPOs remains an open question. Several theoretical models have been proposed, predicting that those may originate either from the Comptonizing corona (Titarchuk & Fiorito 2004) or from the relativistic Lense-Thirring (LT) precession (Stella & Vietri 1998, Ingram et al. 2009). During the outburst rising phase, the observed increase in the QPO frequency has been suggested as evidence for a reduction in coronal size (Chakraborty et al. 2008, Dutta & Chakrabarti 2016). In addition, Zhang et al. (2022) suggested that the evolution of the Type-C QPO frequency is closely linked to changes in the coronal morphology. One of the plausible explanations of LFQPOs is is provided by the two-component advective flow (TCAF) model, which interprets that these oscillations originates from the dynamic behavior of the post-shock region formed within the accretion flow around a BH (Chakrabarti & Titarchuk 1995, Chakrabarti & Manickam 2000, Iyer et al. 2015), which are further supported by numerical simulations and analytical studies (Molteni et al. 1994, Das et al. 2014, Aktar et al. 2018), ascribing the LFQPOs to originate from the post-shock region surrounding the BH which acts as a Comptonizing cloud, upscattering the soft photons to higher energies.

IGR J17091−3624 was discovered in April 2003 by INTEGRAL (Kuulkers et al. 2003). It is a transient BH-XRB that spends most of its lifetime in a dormant state, during which matter gradually accumulates in the accretion disk. When sufficient matter has accumulated, the thermal and viscous instability results in the system to undergo an outburst, which lead to increase in X-ray luminosity by several orders before gradually returning back to dormant state. During an outburst, BH-XRBs evolve through a sequence of accretion states, the LHS, high-soft state (HSS), soft-intermediate state (SIMS) and HIMS, each characterized by distinct spectral and timing properties. This accretion state evolution traces a 'q'-shaped path in the Hardness-Intensity Diagram (HID) (Homan et al. 2001). Following the discovery, IGR J17091−3624 has undergone major outbursts in 2007, 2011, 2016, 2022, and 2025 (Tetarenko et al. 2016; Vincentelli et al. 2025), with each outbursts typically lasting from several weeks to a few months. Throughout these outbursts, the source exhibits the characteristic behavior of transient BH-XRBs, including strong broad-band aperiodic variability, evolving LFQPOs, and transitions between distinct accretion states (e.g., Altamirano et al. 2011; Capitanio et al. 2012; Wang et al. 2024). The 2022 outburst offers an opportunity to investigate the early evolution study, a phase during which the accretion geometry is established and prominent timing features, such as LFQPOs arise.

## 2 Observation and Data Reduction

We analyzed a long-duration AstroSat observation monitored during the onset of 2022 outburst of IGR J17091-3624. The observation commenced on 19 March 2022 at 09:16:41 UT and continued until 22 March 2022 at 00:47:09 UT, corresponding to AstroSat observation ID T05_010T01_9000005016. The lightcurve of the entire observation is shown in Figure 1, it shows all the days covered during the observation period. The lightcurve illustrates the significant variability in the source count rate exhibited throughout the observation. Whereas, Figure 2 presents the corresponding HID for each orbit individually, revealing a segment of the characteristic 'q'-track commonly observed in BH-XRBs during the outbursts.

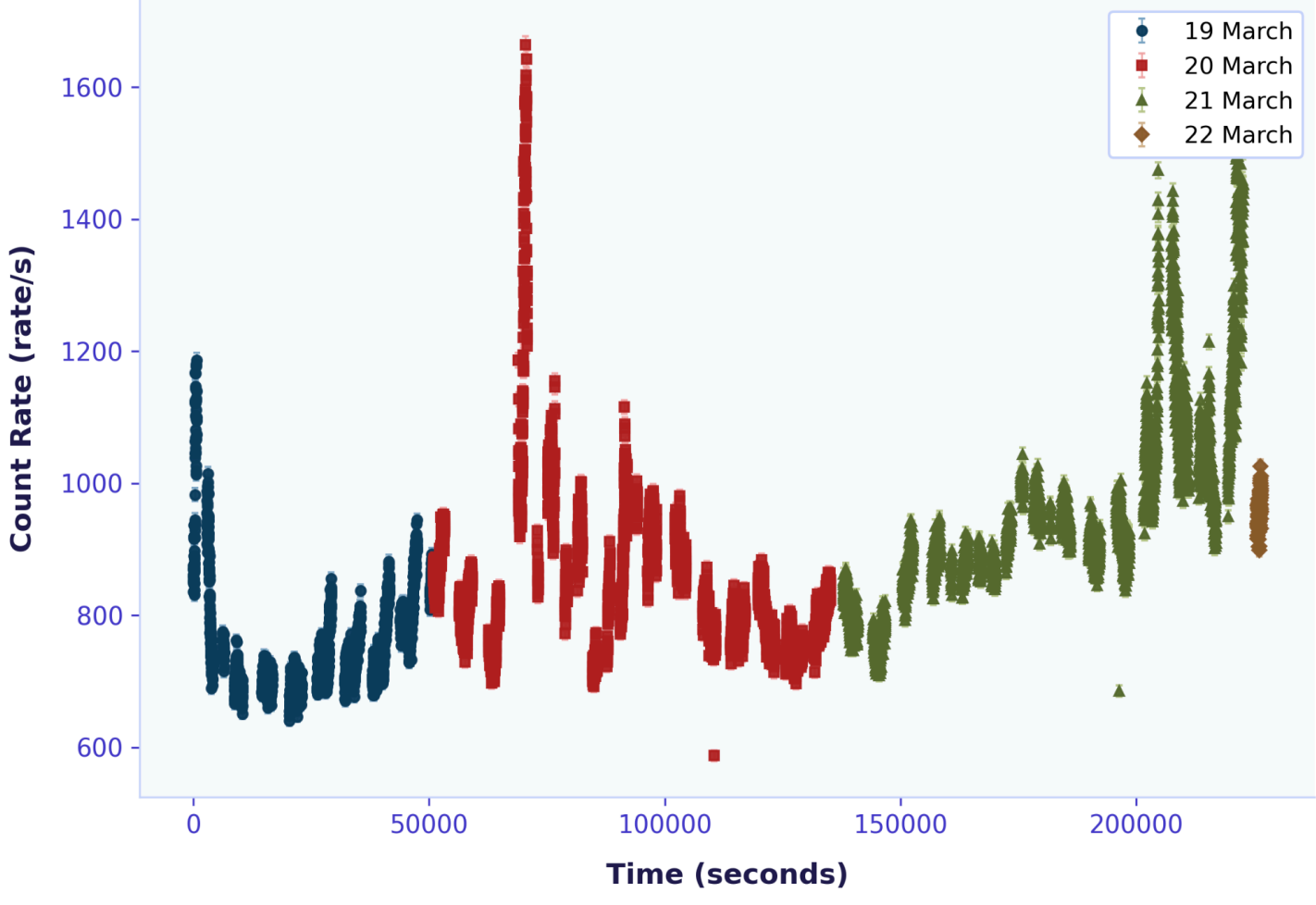

**Figure 1:** LAXPC20 light curve of IGR J17091-3624 in 3-20 keV energy range with 10s timebin. Different markers represent each date covered during the observation in 2022.

AstroSat is India's first dedicated multiwavelength space observatory, carrying five scientific payloads that operate over a wide range of energy. Since this work focuses on X-ray timing analysis, we used data from the Large Area X-ray Proportional Counter (LAXPC), which consists of 3 identical PC units (PCU): PCU10, PCU20, and PCU30. LAXPC operates in the 3-80 keV energy range and provides a fine time resolution of 10 μs with a detector dead time correction of ~42 μs (Yadav et al. 2016, Antia et al. 2017). The LAXPC data was reduced using LAXPCsoftware[1] (version August 15, 2022). The Level-1 orbit files were processed to generate individual Level-2 event files, which were subsequently utilized to extract the required scientific products following the standard analysis pipeline described on the AstroSat Science Support Cell website[2]. In the analysis, we used only the PCU20 data for the analysis, since from March 2018 onward, PCU10 underwent abnormal gain changes, while PCU30 experienced a gas leak.

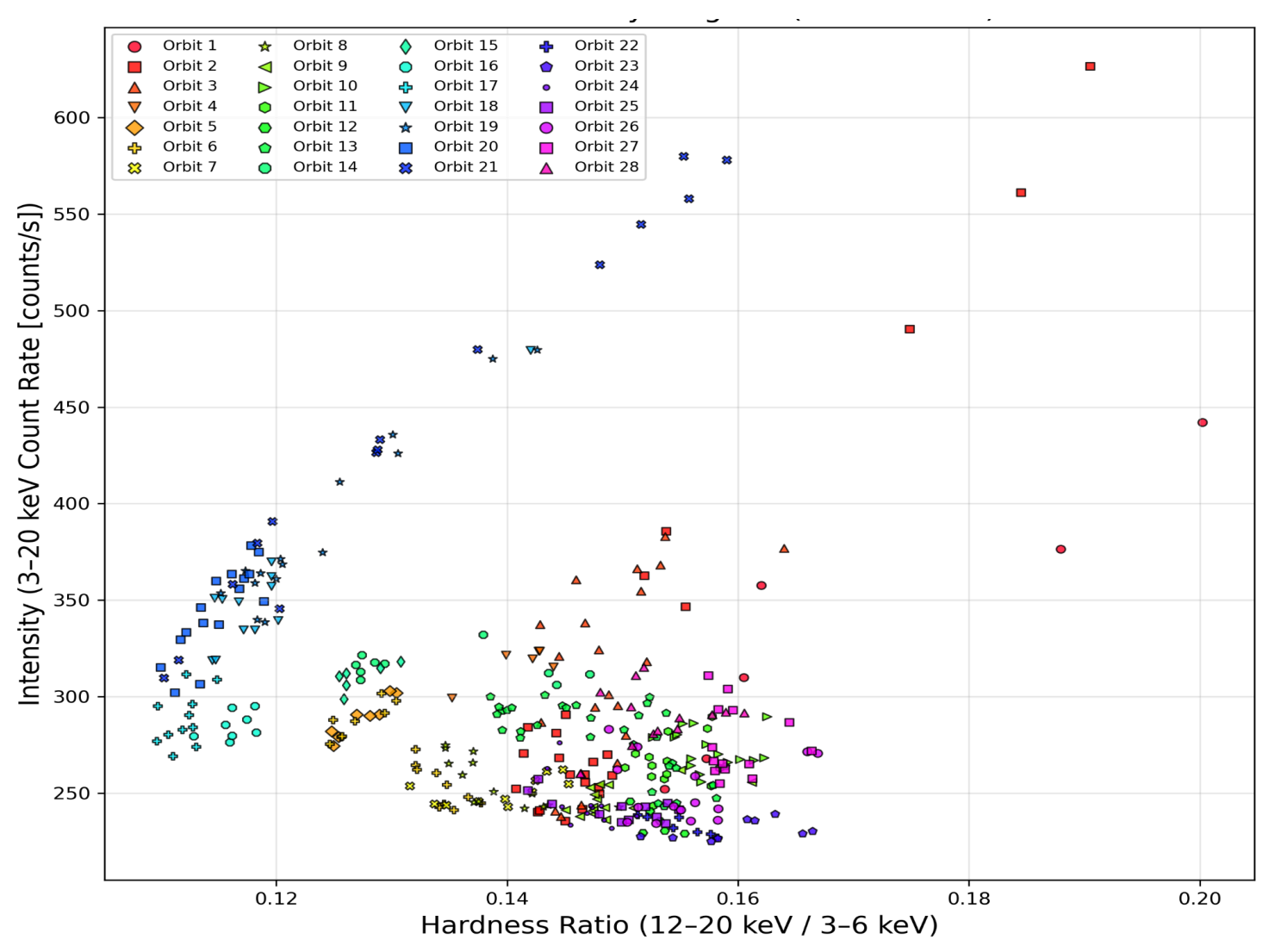

**Figure 2:**Hardness-intensity diagram of IGR J17091−3624 obtained from AstroSat LAXPC20 observations using a time bin of 256 s. Different colors and marker styles denote individual orbits, illustrating the evolution of the source during the observation.

## 3 Timing Analysis

### 3.1 Power Density Spectra

The orbit-wise power density spectra (PDS) of the data has been analyzed and detected Type-

[1] http://astrosat-ssc.iucaa.in/laxpcData
[2] http://astrosat-ssc.iucaa.in/uploads/threadsPageNew_SXT.html

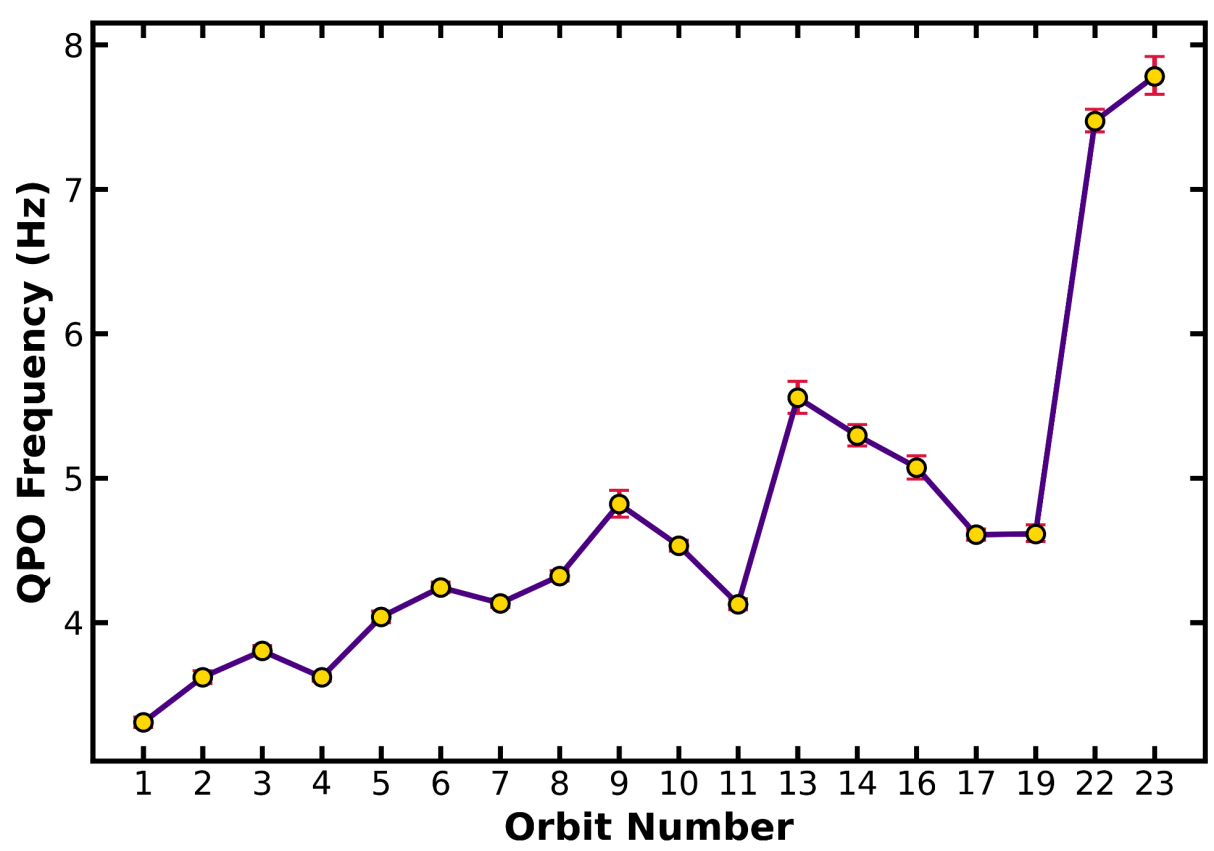

**Figure 3:** Evolution of QPO centroid frequency from 3.30 Hz to 7.78 Hz over time. The X and Y-axes represent the consecutive orbits and the QPO frequency (Hz), respectively.

**Table 1:** Best-fit Lorentzian parameters of the detected Type-C QPOs for different orbits.

| **Orbit** | **QPO Frequency (Hz)** | **Width (Hz)** | **Norm ($\times 10^{-3}$)** | **Q-factor** | **QPO significance ($\sigma$)** |
|---|---|---|---|---|---|
| 1 | 3.30 ± 0.02 | 0.37 ± 0.06 | 1.57 ± 0.19 | 8.77 ± 2.16 | 6.14 |
| 2 | 3.62 ± 0.03 | 0.73 ± 0.08 | 3.63 ± 0.29 | 4.98 ± 1.02 | 5.59 |
| 3 | 3.80 ± 0.02 | 0.47 ± 0.07 | 2.43 ± 0.24 | 7.98 ± 1.80 | 6.42 |
| 4 | 3.62± 0.02 | 0.47 ± 0.06 | 2.83 ± 0.28 | 7.65 ± 1.64 | 6.40 |
| 5 | 4.04 ± 0.02 | 0.72 ± 0.08 | 2.76 ± 0.25 | 5.63 ± 0.98 | 7.67 |
| 6 | 4.24 ± 0.02 | 0.81 ± 0.07 | 3.05 ± 0.20 | 5.23 ± 0.82 | 8.78 |
| 7 | 4.13 ± 0.02 | 0.63 ± 0.07 | 2.31 ± 0.24 | 6.60 ± 1.21 | 6.50 |
| 8 | 4.32 ± 0.02 | 0.56 ± 0.07 | 1.56 ± 0.15 | 7.69 ± 1.69 | 6.56 |
| 9 | 4.82 ± 0.05 | 1.13 ± 0.19 | 1.49 ± 0.22 | 4.26 ± 0.96 | 5.14 |
| 10 | 4.53 ± 0.02 | 0.72 ± 0.08 | 1.38 ± 0.12 | 6.29 ± 1.10 | 7.52 |
| 11 | 4.13 ± 0.02 | 0.87 ± 0.07 | 2.09 ± 0.14 | 4.74 ± 0.65 | 8.31 |
| 13 | 5.55 ± 0.06 | 1.00 ± 0.19 | 1.52 ± 0.18 | 5.55 ± 2.08 | 3.76 |
| 14 | 5.30 ± 0.04 | 1.09 ± 0.16 | 1.49 ± 0.18 | 4.84 ± 1.20 | 5.44 |
| 16 | 5.07 ± 0.05 | 1.06 ± 0.17 | 1.63 ± 0.23 | 4.80 ± 1.26 | 4.88 |
| 17 | 4.61 ± 0.02 | 0.68 ± 0.08 | 2.01 ± 0.17 | 6.76 ± 1.20 | 7.65 |
| 19 | 4.61 ± 0.03 | 1.12 ± 0.12 | 2.75 ± 0.26 | 4.12 ± 0.98 | 3.53 |
| 22 | 7.47 ± 0.05 | 0.88 ± 0.14 | 0.55 ± 0.06 | 8.43 ± 2.46 | 5.68 |
| 23 | 7.78 ± 0.07 | 1.18 ± 0.23 | 0.41 ±0.06 | 6.55 ± 2.27 | 4.53 |

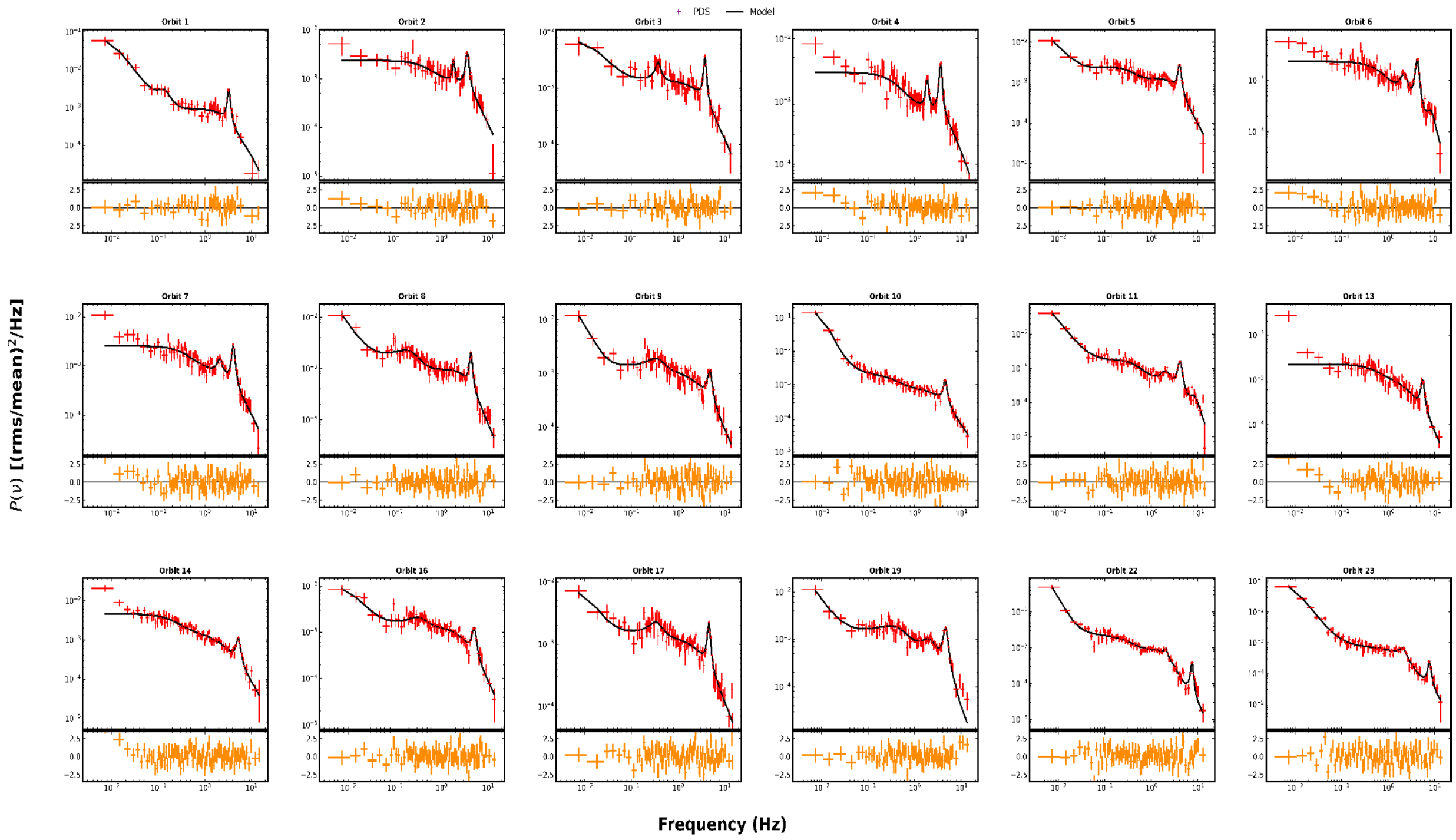

**Figure 4:** PDS of IGR J17091-3624 for the subsequent orbits within the 3-20 keV range using PCU20. The subplots show a clear presence of QPO features in the observation.

C QPOs whose centroid frequency evolves throughout the observation. The observed QPOs are identified as Type-C based on the presence of a prominent QPO superimposed on the characteristic flat-topped broad-band noise continuum, often accompanied by harmonic features. Of the 23 orbits, 18 exhibit these canonical Type-C timing characteristics. It should be noted that some adjacent orbits lying in the same state were merged to improve the statistical quality of the PDS, thereby enhancing the significance of the detected timing features. The best-fitting parameters, including the QPO centroid frequency ($\nu$), full width half maximum (FWHM) and its normalization (N) alongwith estimated Q factor, and QPO significance ($\sigma = N/N_{low_err}$), where N and $N_{low_err}$ denote the Lorentzian normalization and its lower error bar, respectively, are listed in Table 1. The LAXPC PDS were generated using the subroutine *laxpc_find_freqlag* of the LAXPCsoftware package. Significant narrow QPO features were detected to evolve in the frequency range of 3.30-7.78 Hz. The evolution of the QPO centroid frequency over successive orbits is shown in Figure 3. To probe the evolution of the timing properties, light curves with a time resolution of 33.3 ms were divided into segments of ~136.4 s, and the PDS from individual orbits were generated. Each PDS was fitted with a multiple-Lorentzian model consisting of one zero-centered Lorentzian to describe the broad-band noise and additional Lorentzian components to model the QPO and its harmonic, when present (Belloni et al. 2002). The best-fitted PDS for the orbits showing prominent QPO features are presented in Figure 4.

### 3.2 Time Lag and rms calculation

The subroutine *laxpc_find_freqlag* enables us to compute the energy-dependent time lag and frms spectra at the QPO centroid frequency. The calculations were performed over a frequency interval of $\nu$-$\Delta\nu$ to $\nu$+$\Delta\nu$, where $\nu$ is the QPO centroid frequency and $\Delta\nu$ is the half-width of the selected frequency range. The energy range of 3.0–22.0 keV was divided into several energy bins, and the time lags were calculated with respect to the 3.0–4.0 keV reference energy band. The resulting energy-dependent time lag and frms spectra are presented in Figures 5 and 6, respectively. In most of the orbits, it is observed that the soft photons are lagging behind the hard photons whereas we obtain a positive correlation of the frms with the energy.

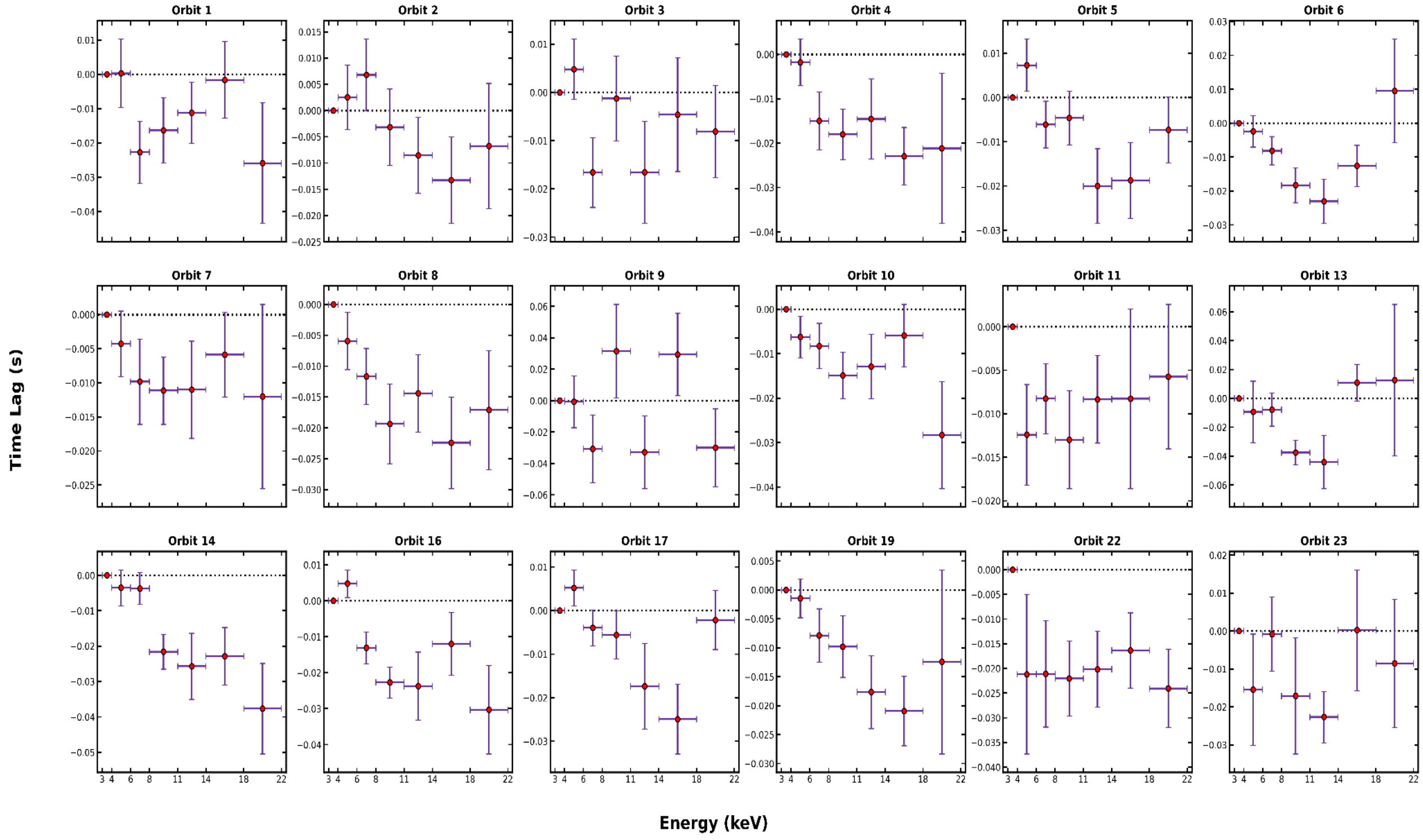

**Figure 5:** Energy dependent time-lag spectra at the QPO centroid frequency for different orbits of IGR J17091−3624 as a function of photon energy.

## 4 Result and Discussion

The timing analysis of the AstroSat/LAXPC observations during the onset phase of the 2022 outburst of IGR J17091-3624 reveals significant Type-C LFQPOs in the observation. The QPO centroid frequency evolves from 3.30 ± 0.02 Hz to 7.78 ± 0.03 Hz during the observation, indicating changes in the geometry of the inner accretion flow. The detected QPOs have Q factors in the range Q ≈ 4.12 ± 0.98 to 8.77 ± 2.16 and detec-

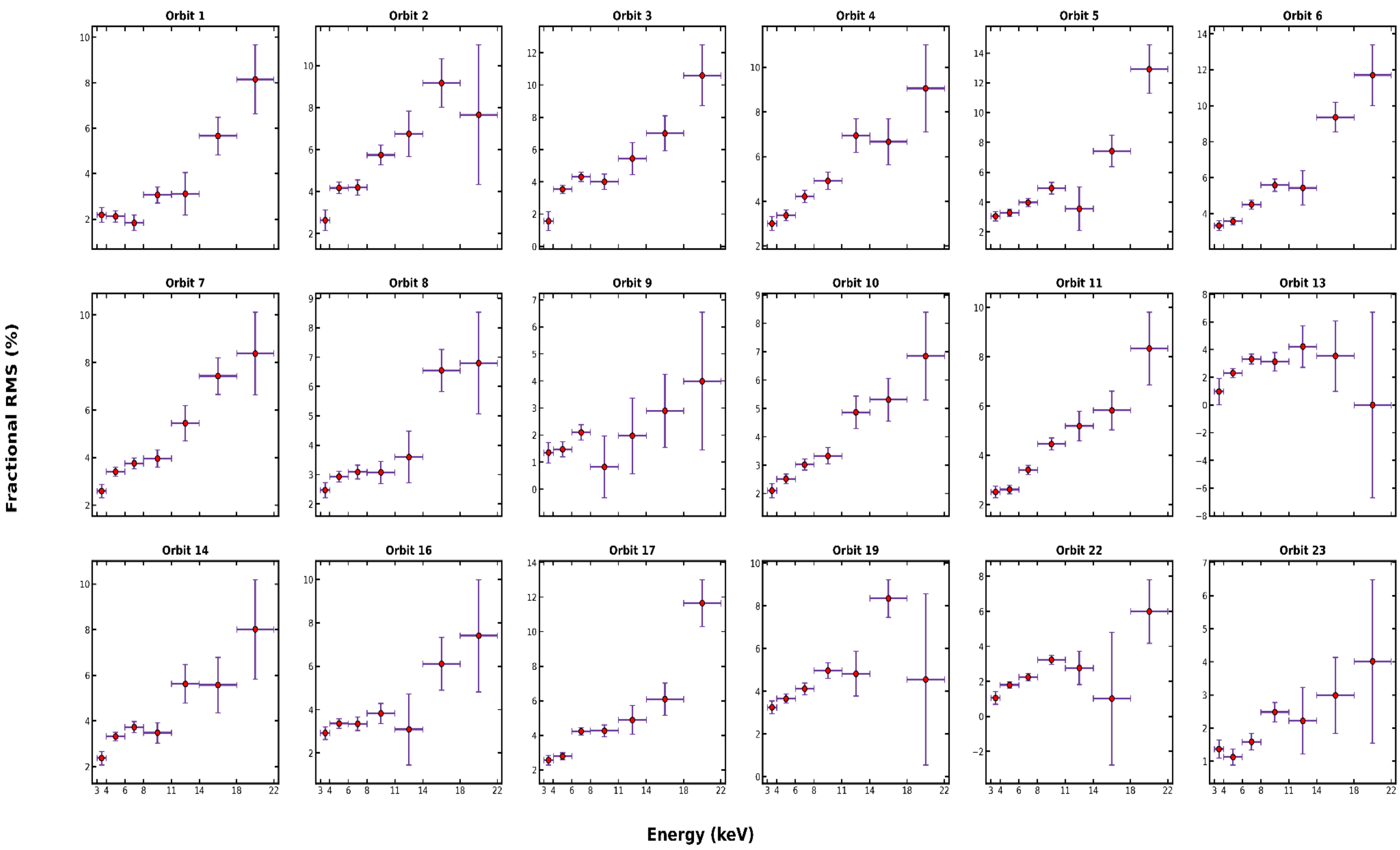

**Figure 6:** Energy-dependent frms spectra at the QPO centroid frequency for different orbits of IGR J17091−3624 as a function of photon energy.

-tion significance is >3σ for all, which varies between 3.53σ and 8.78σ, confirming that the observed features are statistically robust. The PDS are characterized by narrow QPO peaks superimposed on strong broad-band noise, with harmonic features detected in several orbital segments. According to the observational classification of LFQPOs (Casella et al. 2005), the combination of centroid frequencies of a few hertz, significant narrow peaks, strong broad-band noise, and harmonics identifies these oscillations as Type-C QPOs. The integrated QPO frms amplitude varies between 2.0% and 6.0%. Such rms amplitudes are typical of Type-C QPOs observed during the HIMS, where the increasing contribution of the thermal disk reduces the overall variability relative to the LHS (Casella et al. 2004; Belloni et al. 2011). Broadly consistent with this, the energy-dependent rms spectra show a systematic increase with photon energy in nearly all orbital segments, rising from approximately 1–3% below 5 keV to 8–12% above 15 keV (see Figure 6). This trend indicates that the QPO variability is dominated by the hard Comptonized emission from the inner hot flow, while the comparatively stable thermal disk emission dilutes the variability at lower energies (Sobolewska & Życki 2006). The energy-dependent time-lag spectra exhibit soft lags in most of the orbits, with delay amplitudes of approximately 5–30 ms, although the highest-energy bins are associated with larger uncertainties because of limited photon statistics (see Figure 5). The soft lags suggest that variations in the hard X-ray emission generally precede those in the softer bands. Such lag behavior may result from thermal reverberation of hard photons by the accretion disk, spectral evolution of the Comptonized emission, or changes in the illumination geometry of the inner accretion flow (Uttley et al. 2014). Since these mechanisms can produce similar lag signatures, and also due to large uncertainties, the observed time lags alone cannot uniquely determine the physical origin of the QPO.

The AstroSat observation overlaps with the HIMS reported by Wang et al. (2024) and extends close to the HIMS-SIMS transition. The observed increase in the Type-C QPO frequency and the evolution of the timing properties are consistent with this spectral state evolution. The increase in QPO frequency and the increasing rms toward higher energies, suggests that the characteristic radius of the hot inner flow gradually decreases as the outburst evolves, the decrease of the amplitude of rms% towards the higher QPO centroid frequency with the progress of time is also consistent the scenario of evolution of the source through HIMS, where the increasing contribution of the thermal disk progressively reduces the overall variability. The strong energy dependence of the rms further supports an origin of the oscillation within the Comptonizing region. These timing properties are also consistent with the LT precession model (Stella & Vietri 1998, Ingram et al. 2009, Ingram & Done 2011), in which the precession of the inner hot flow modulates the Comptonized emission and gives rise to the observed Type-C QPOs. Although the present timing analysis cannot uniquely distinguish among different theoretical models, the collective observational evidence, namely the QPO frequency evolution, statistically significant detections, broad-band noise, harmonic structure, moderate integrated rms, increasing rms with energy, and predominantly soft energy-dependent lags, are consistent with an origin in the Comptonizing inner flow during the HIMS.

## 5 Conclusion

We investigated the rapid X-ray timing properties of IGR J17091−3624 during the early phase of its 2022 outburst using AstroSat/LAXPC observations. Significant LFQPOs were detected in 18 orbital segments, with centroid frequencies evolving from 3.30 to 7.78 Hz. The QPOs exhibit moderate Q factors (Q ≈ 4-9), detection significance above 3σ, strong broad-band noise, harmonic features in several observations, and integrated frms amplitudes of 2–6%, consistent with the observational characteristics of Type-C QPOs. There is a positive correlation between the energy-dependent frms with photon energy, indicating that the rapid variability is dominated by the Comptonizing inner flow. The energy-dependent time-lag spectra show hints of soft lags of 5–30 ms, suggesting a close relation between the Comptonizing region and the accretion disk through processes such as disk reprocessing. Combined with the HIMS classification reported by Wang et al. (2022) during this period and the observed increase in QPO frequency, these results support a scenario in which the hot inner flow evolves inward while remaining the primary site of rapid X-ray timing variability. Overall, the observed timing properties are broadly consistent with the truncated accretion disk geometry and the LT precession interpretation for the origin of Type-C QPOs in BH-XRBs (Done et al. 2007, Ingram et al. 2009, Ingram & Done 2011).

Though the present work focuses on the X-ray timing properties of IGR J17091−3624, a simultaneous broadband spectral analysis will be carried out in future to directly correlate the evolution of the timing properties with changes in the accretion flow, including the physical properties and geometry of the accretion disk and Comptonizing corona. Another limitation of the present study is the availability of data covering only the early phase of the 2022 outburst. Continuous observations covering the entire outburst evolution would enable a comprehensive investigation of the evolution of both timing and spectral characteristics across different accretion states, thereby providing stronger constraints on the physical origin of QPOs and the evolution of geometry of the inner accretion flow.

**Acknowledgements**: This research work has made use of the software developed and distributed by the High Energy Astrophysics Science Archive Research Center (HEASARC). The study is based on observations obtained by LAXPC onboard the AstroSat mission of the Indian Space Research Organisation (ISRO). We thank the LAXPC Payload Operations Centre (POC) at the Tata Institute of Fundamental Research (TIFR), Mumbai, for providing the data through the Indian Space Science Data Centre (ISSDC) archive and for making the required software tools available. S.B. acknowledges the financial support provided by the INSPIRE Fellowship (Grant No. DST/INSPIRE Fellowship/IF220164) from the Department of Science and Technology (DST), Ministry of Science and Technology, Government of India. The Author B.S. acknowledges the IUCAA Visiting Associateship Program.